\documentclass[%
 reprint,
 amsmath,amssymb,
 aps,
 prb,
floatfix,
]{revtex4-1}

\usepackage{graphicx}
\usepackage{dcolumn}
\usepackage{bm}
\usepackage{color}
\usepackage{times}
\usepackage{amsmath}
\usepackage{float}
\newcommand{\beginsupplement}{%
	\setcounter{table}{0}
	\renewcommand{\thetable}{S\arabic{table}}%
	\setcounter{figure}{0}
	\renewcommand{\thefigure}{S\arabic{figure}}%
}

\begin{document}

%
\title{
Harnessing asymmetry to  reprogram nonlinear metamaterials on-the-fly with no moving parts}
\author{ Majid Kheybari and Osama R. Bilal}
\affiliation{Department of Mechanical engineering, University of Connecticut, Storrs, USA}

\date{\today}
\begin{abstract}
Various two-dimensional fabrication methods, such as deposition, etching, milling, laser cutting, and water jetting,  suffer from asymmetry between the top and the bottom surface of fabricated parts. Such asymmetry is usually undesirable and  can compromise functionality, or at least add uncertainty to  fabricated components. The common practice is to assume symmetry between the top and the bottom surfaces by using average dimensions. In this study, we harness such asymmetry to realize metamaterials with dynamically tunable (i. e., re-programmable) properties. Our metamaterial is composed of identical unit cells with two concentric Archimedean spiral cuts and a permanent magnet embedded in the unit cell's center. By utilizing external electromagnets, we further amplify the fabrication asymmetry, through the inherent asymmetry between a repulsive vs attractive state between the permanent magnets and the electromagnets. We demonstrate the utility of our metamaterials by programming its spatiotemporal response in both time and frequency even in the presence of high amplitude harmonic excitation. Our findings can be utilized for broad range of applications, from seismic sensing at low frequency to ultrasonic imaging at higher frequencies.

\end{abstract}

\maketitle

Metamaterials are arrangements of basic building blocks, or unit cells, that repeat in space giving rise to  unconventional negative effective properties \cite{lu2009phononic,maldovan2013sound, hussein2014dynamics, cummer2016controlling}. For example, metamaterials can support the formation of frequency band gaps where elastic/acoustic waves are forbidden from propagation \cite{bilal2013trampoline,deymier2013acoustic}. Potential applications of such metamaterials are sound insulation/attenuation \cite{yang2010acoustic, mei2012dark}, wave-guiding\cite{ding2010compact, li2019active,wang2023reconfigurable}, focusing \cite{torrent2007acoustic, yang2004focusing, li2012acoustic}, cloaking \cite{cummer2007one, torrent2008acoustic, cummer2008scattering}, filtering \cite{pennec2004tunable, zhu2013metamaterial, liao2021acoustic} or localization\cite{boechler2010discrete}. Despite their great potential in manipulating waves, metamaterials with fixed properties could be less desirable in practical applications. Therefore, a great deal of research is dedicated to realizing metamaterials with tunable properties. Examples of tunable acoustic/elastic metamaterials include utilizing varying air pressure \cite{lee2012highly, langfeldt2016membrane}, external load \cite{wang2014harnessing, kheybari2022tunable,roshdy2023tunable}, buckling  \cite{zhang2022buckling,kheybari2022programmability}, piezoelectric patches   \cite{airoldi2011design, sugino2017investigation, chen2013tunable, casadei2012piezoelectric, hou2015tunable, sugino2020digitally, yi2020programmable, celli2018pathway, alshaqaq2022programmable,degraeve2014bragg,li2022actively}, electric fields  \cite{xiao2015active}, electromagnets\cite{ma2018shaping, wang2016tunable, chen2016tunable, yang2017programmable, haghpanah2016programmable,salari2019negative, liu2019designing, gao2023investigation}, magnetic fields \cite{bilal2017bistable,bilal2017reprogrammable,robillard2009tunable, vasseur2011band, allein2016tunable, chen2014active, zhao2017membrane,guell2020magnetoelastic,zhang2021tunable}, acoustic trapping \cite{caleap2014acoustically}, granular contacts \cite{li2014granular, boechler2011bifurcation}, and shape memory effects \cite{chuang2019bandgap,hu2018dynamic,candido2018tunable,lv2020shape,gliozzi2020tunable, li20214d,sepehri2021manipulation,song2022smoothly}. Some of these tuning methods are irreversible, relatively slow, require manual intervention, moving parts, direct contact with the metamaterials or limited to low amplitude waves \cite{cummer2016controlling, chen2018review, zangeneh2019active, kumar2019recent, ji2021recent, xia2022responsive}.  


\begin{figure}[!h]
\centering
\includegraphics{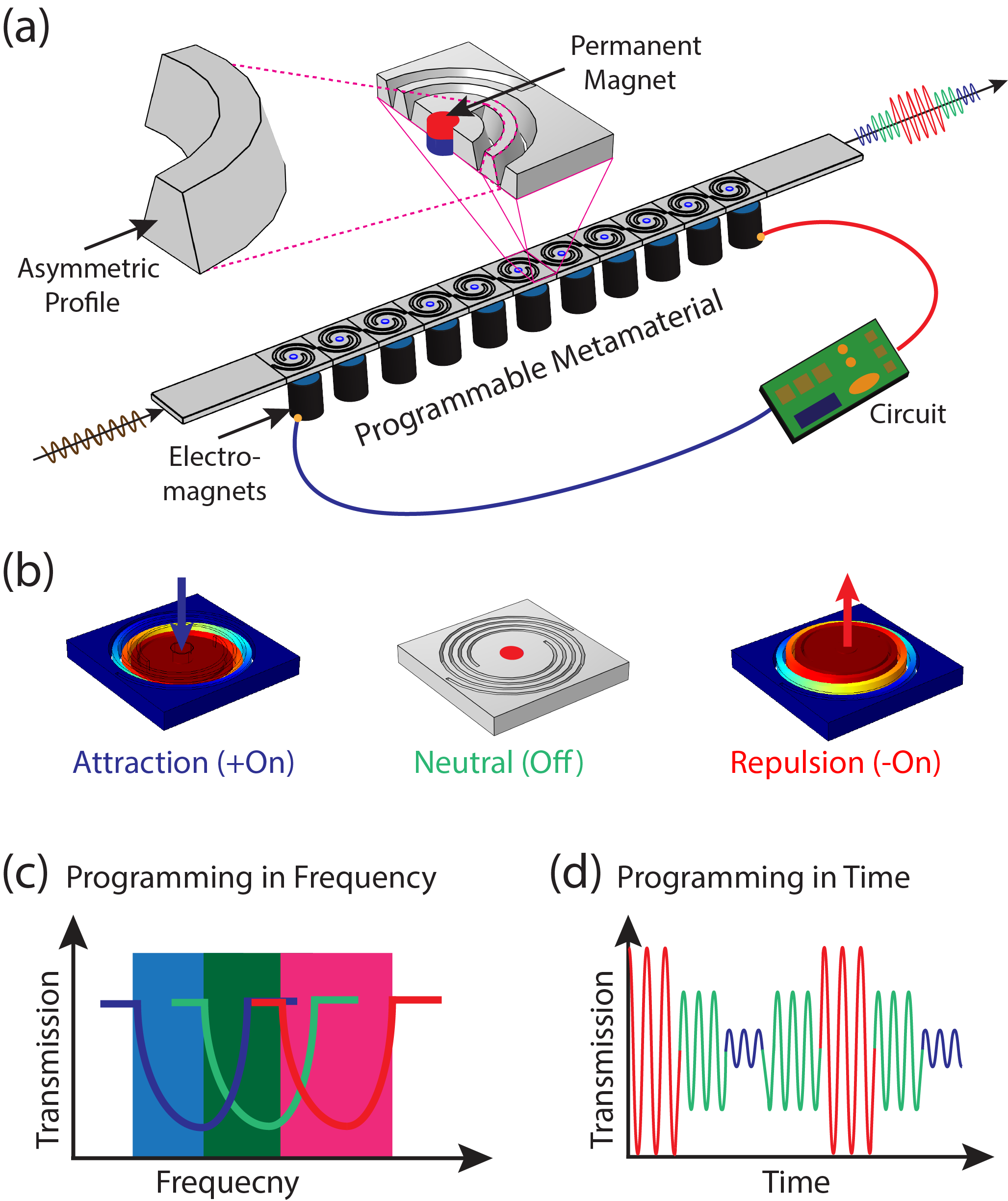}
\caption{\label{fig:Figure-1}\textbf{Concept: }(a) Schematic of an experiment setup. (b) The metamaterial can have one of three states depending on the programmed polarity of the electromagnets: (i) attraction between the permanent magnets and the electromagnets, (ii) neutral when the electromagnets are off, or (iii) repulsion between the permanent magnets and the electromagnets. (c-d) The metamaterial response can be tuned in both the frequency and time domain.}
\end{figure}

In this study, we design, analyze and experimentally realize a class of re-programmable metamaterial that can be actively tuned to manipulate elastic waves \textit{on-the-fly} with \textit{no contact, no moving parts, continuous tunability}, while being able to filter targeted fundamental or higher harmonics, by design. We refer to our metamaterial as re-programmable, because the tuning of the material's spatiotemporal properties is reversible. Our metamaterial is composed of a basic building block that repeats periodically in space (i.e., unit cell). Each unit cell features two concentric Archimedean spiral cuts with a permanent magnet embedded at its center. All the permanent magnets are oriented in the same direction (i.e., north facing up). The spiraling cuts within the unit cell have asymmetric profiles across the thickness  (Fig.\ref{fig:Figure-1} (a)). To affect a dynamical change in our material's properties, we place an array of electromagnets underneath the metamaterial sample. The electromagnets are connected in parallel to a DC power supply and are controlled by a pre-programmed electric circuit. The metamaterial can have one of three states depending on the programmed polarity (positive, off, or negative) of the electromagnets: (i) attraction between the permanent magnets and the electromagnets, (ii) neutral when the electromagnets are off, or (iii) repulsion between the permanent magnets and the electromagnets (Fig.\ref{fig:Figure-1} (b)). The spatiotemporal characteristics of the metamaterial are dynamically tuned by programming the state of the metamaterial's surrounding magnetic field over time. The tunability of our metamaterial can be achieved for a band of frequencies by shifting the band gap region within the frequency spectrum (Fig.\ref{fig:Figure-1} (c)). The tunability can also be  achieved for the frequency contents of a single harmonic excitation  over specific time intervals by selectively attenuating particular harmonics within the transmitted signal (Fig.\ref{fig:Figure-1} (d)).

\begin{figure}[t]
\centering
\includegraphics{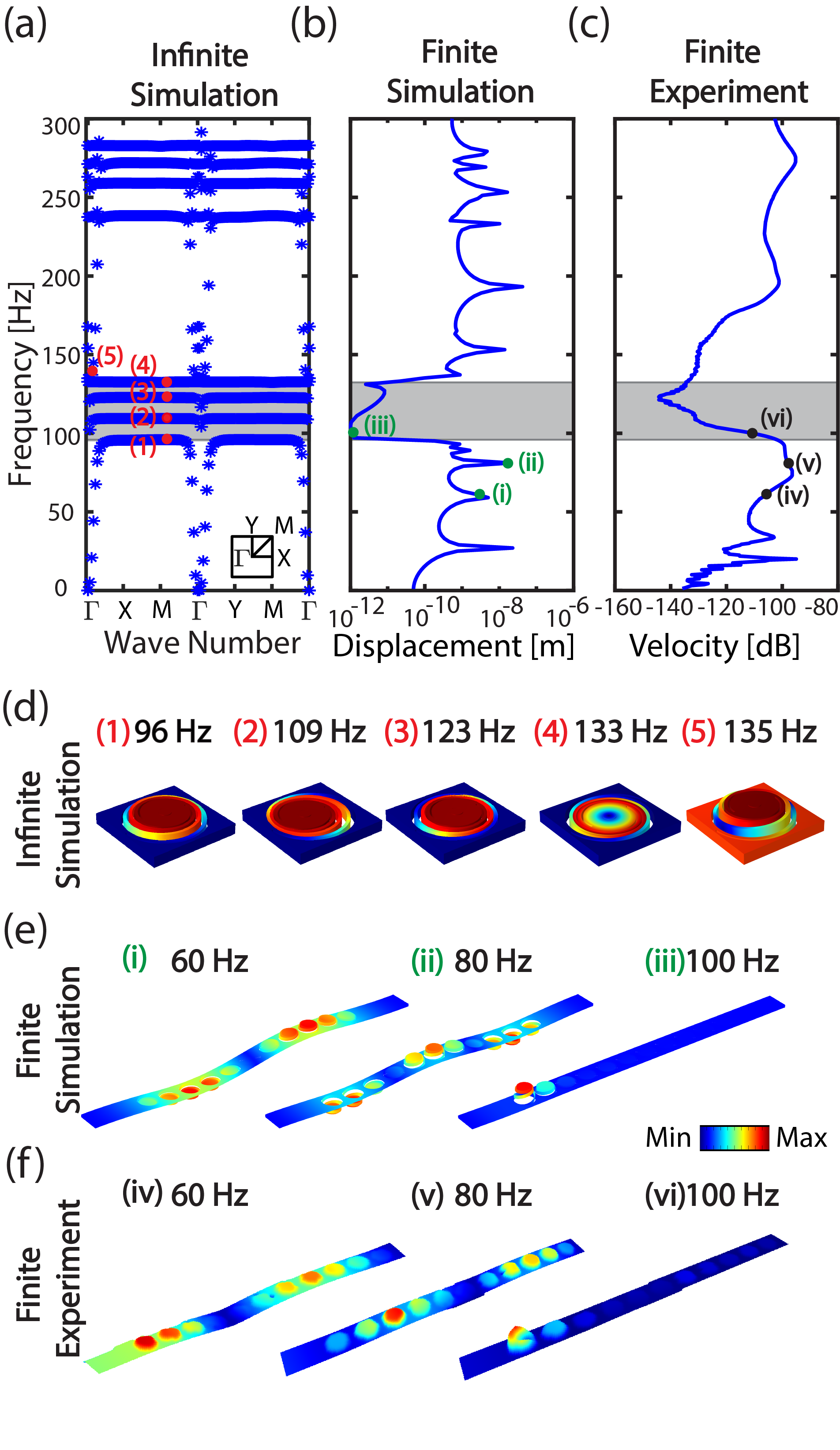}
\caption{\label{fig:Figure-2}\textbf{Unit cell analysis:} (a) Numerically computed dispersion curves for infinitely repeated symmetric unit cell. (b) Numerically computed frequency response function of the transmission in the metamaterial having an array of unit cells of size 11x1. (c) Experimentally measured frequency response function of the transmission in the metamaterial having an array of unit cells of size 11x1. (The gray shaded region highlights the frequency ranges of the band gaps). (d) Selected mode shapes of the numerically computed dispersion curves. (e) Selected mode shapes of numerically computed frequency response function of the transmission in a finite structure. (f) Selected mode shapes of experimentally measured frequency response function of the transmission in a finite structure.}

\end{figure}

To study the dynamics of our metamaterial, we consider a single unit cell with two concentric Archimedean spiral cuts and an embedded cylindrical permanent magnet at its center. We assume an infinite repetition of the unit cell in space by implementing a Bloch solution \cite{deymier2013acoustic} in the form: $\textbf{u}(\textbf{x},\kappa;t)=\tilde{\textbf{u}}(\textbf{x},\kappa)e^{i(\kappa.\textbf{x}-\omega t)}$, where  $\tilde{\textbf{u}}$ is the Bloch displacement vector,   $\textbf{x}$ is the position vector, $\kappa$ is the wave number, $\omega$ is the frequency, and $t$ is time. By employing the Bloch solution, we write the dispersion relation as an eigenvalue problem in the form: $ [-\omega^2\textbf{M}+\textbf{K}(\kappa)] \boldsymbol{u} = 0$, where $\textbf{M}$, $\textbf{K}$ are the mass and stiffness matrices, respectively. We obtain the dispersion curves, correlating frequency and wavenumber, for our unit cell by solving the formulated eigenvalue problem. We use the finite element method to calculate the dispersion curves (Fig. \ref{fig:Figure-2} (a)) utilizing COMSOL multi-physics version 6.0. The dispersion curves show a complete (i.e., in all wave propagation directions) band gap between 96 Hz and 135 Hz, or 34\% band gap (gray highlighted region in Figure \ref{fig:Figure-2} (a-c)). The mode shapes at the edges of the band gap correspond to out-of-plane displacement profiles for the center of the unit cell (Fig.\ref{fig:Figure-2} (d)). Within the highlighted band gap frequency range, there exist three transmission branches (at 109, 123, and 133 Hz) corresponding to in-plane modes (Fig.\ref{fig:Figure-2} (d)). The position of the band gap within the frequency spectrum is a function of the unit cell design parameters \cite{foehr2018spiral,bilal2020enhancement,kheybari2022programmability}. For example, decreasing the  width of the spiral cuts, while keeping all other parameters fixed, increases the band-gap's central frequency (Fig.\ref{fig:SFigure-2}). Such decrease in the width of the cut, increases the thickness of the spiraling arms, which in turn increase the arms' stiffness and vice-versa. However, such change in unit cell parameters is permanent, i.e., once fabricated, the unit cell design can not be altered. An alternative route to achieve tunability of the band gap frequency range is to selectively add dynamically controllable stiffness to the modes at the edge of the band gap. Given the displacement profile of the mode shapes at the edge of the gap, an added stiffness to the center of the spirals, changes the frequency of the mode and therefore shifts the edge of the band gap. We achieve such shift through programming the surrounding magnetic field both in time and space. 

 
To verify our band structure calculations, using the infinite  unit cell model (i.e., with Bloch periodic boundary conditions), we numerically simulate a meta-structure composed of a finite number of unit cells. We consider an array of 11 unit cells with fixed-fixed boundary conditions. We apply an out-of-plane harmonic excitation at one end of the metamaterial and record the transmission of the wave at the other end of the metamaterial. The result shows a clear attenuation through the metamaterial at the predicted band gap region as highlighted in gray (Fig.\ref{fig:Figure-2} (b)). In addition, we examine three vibrational modes of the simulated finite structure inside and outside the band gap frequency range (Fig.\ref{fig:Figure-2} (e)). The first two modes at 60 and 80 Hz show clear transmission from the left-end to the right-end within the metamaterial (Fig.\ref{fig:Figure-2} (e)i and ii), while the third mode at 100 Hz shows clear localization of the excited wave on the left end of the metamaterial(Fig.\ref{fig:Figure-2} (e)iii).

\begin{figure}
\centering
\includegraphics{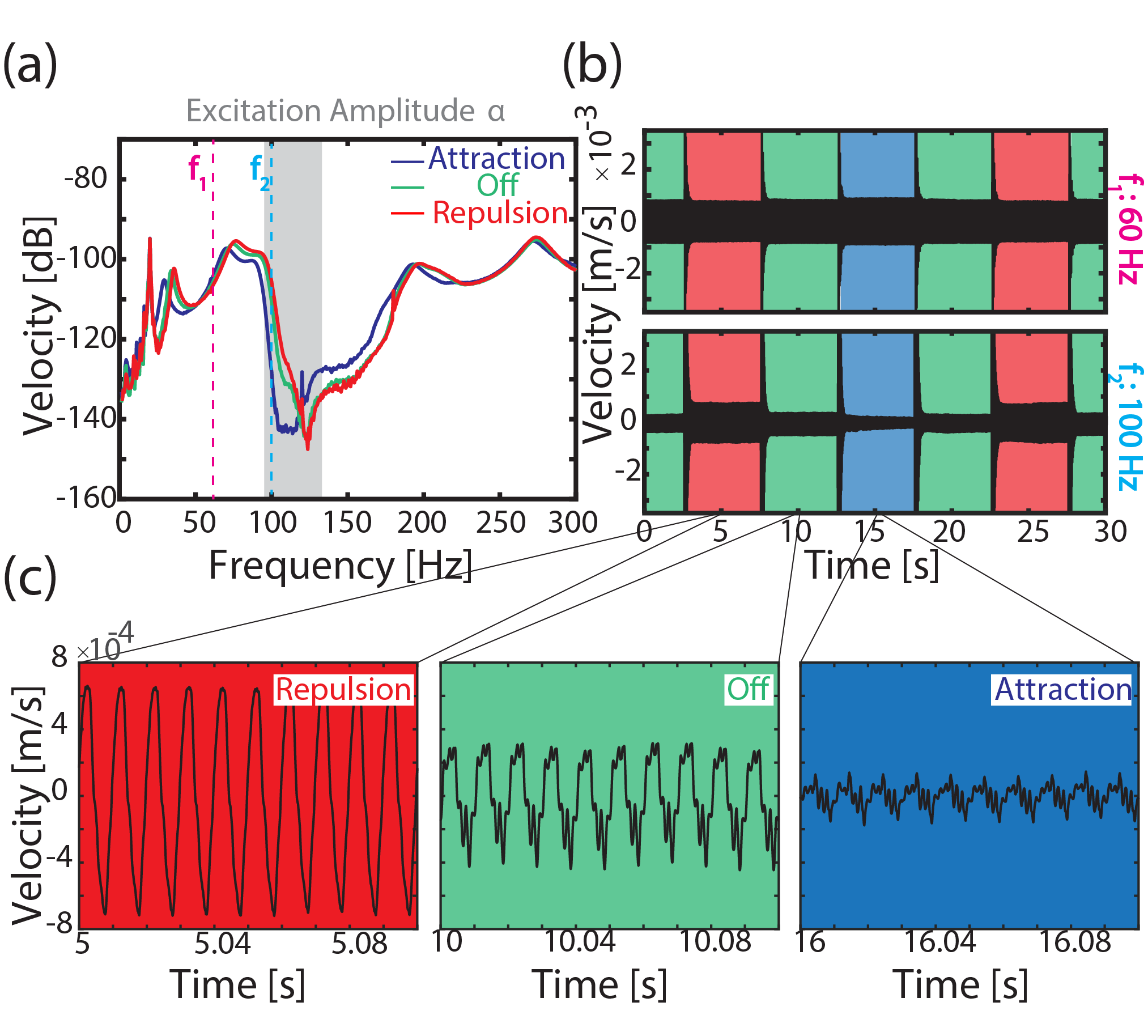}
\caption{\label{fig:Figure-3}\textbf{Programming metamaterial in frequency:} (a) Experimentally measured frequency response of the proposed metamaterial under excitation amplitude of $\alpha$. (b) Time-signal of selected frequencies $f_2$ within the band gap,  and $f_1$  within the pass band (aqua and red dashed lines in panel a). (c) Zoomed in window of the time-signal for repulsion (highlighted in red), off (highlighted in green) and attraction (highlighted in blue).}
\end{figure}

To experimentally validate the numerical simulations of both the infinite and the finite structures, we fabricate a physical prototype of our metamaterial consisting of an array of 11x1 unit cells. We replicate the finite meta-structure simulations experimentally, by harmonically exciting the metamaterial at one-end with a mechanical shaker. We measure the amplitude of the response at the other end of the meta-structure,  using a scanning laser vibrometer, as we sweep through different excitation frequencies. The resulting frequency response function (FRF) shows a good agreement with our numerical simulations, particularly at the attenuated band gap region (Fig.\ref{fig:Figure-2} (c)). Furthermore, we experimentally scan the entire meta-structure at three different excitation frequencies, to visualize the transmission of the wave from the left-end to the right-end. Both scans at 60 and 80 Hz show clear transmission of the wave through the meta-structure from left to right. The third experimental scan at 100 Hz, however, shows a clear localization of the wave at the left-end of the meta-structure (Fig.\ref{fig:Figure-2} (f); Movies S1-S3).

Once fabricated, metamaterial properties are usually fixed. Having the ability to tune their properties in real-time without moving parts could be invaluable for future applications. To test our proposed concept of dynamic tunability in the frequency domain, we harmonically excite the metamaterials in the presence of two different magnetic-field polarizations and record the amplitude of the transmitted elastic waves. We superimpose the measured steady state responses of the metamaterial (FRF) with the electromagnets ON and OFF. We consider both cases when the magnetic field is ON and in repulsion or ON and in attraction relative to the unit cells' permanent magnets (Fig. \ref{fig:Figure-3} (a)). There exists a shift of the band gap frequency due to the presence of the magnetic field. More importantly, we observe a clear difference between the metamaterial's steady state response in repulsion vs attraction. This difference stems from the inherent asymmetry in the magnetic field as the core of the unit cell is pushed closer or farther away from the electromagnets, which is amplified by the designed geometric asymmetry (top vs bottom) within the unit cell.

In order to characterize the behavior of the metamaterial in the time domain, we keep the harmonic excitation source fixed to a single amplitude ($\alpha$). We then record the transmitted elastic wave through the metamaterial under different magnetic fields for two frequencies $f_1$ = 60 Hz and $f_2$ = 100 Hz. We program the magnetic field to change between OFF, ON-attraction and ON-repulsion every 5 seconds (Fig.\ref{fig:Figure-3} (b)). On one hand, at $f_1 = 60$, the transmission amplitude is almost constant despite the change in the magnetic field, because $f_1$ is a pass band frequency. On the other hand, at $f_2 = 100$, the transmission is roughly 6 times higher in the case of repulsion than attraction, with the OFF case having an intermediate transmission amplitude (Fig.\ref{fig:Figure-3} (b)). For both frequencies, $f_1 \& f_2$, the change between transmission amplitudes, or the transience, takes $\approx 0.2~sec$ (Fig \ref{fig:SFigure-9}; Movies S6).

\begin{figure}[b]
\includegraphics{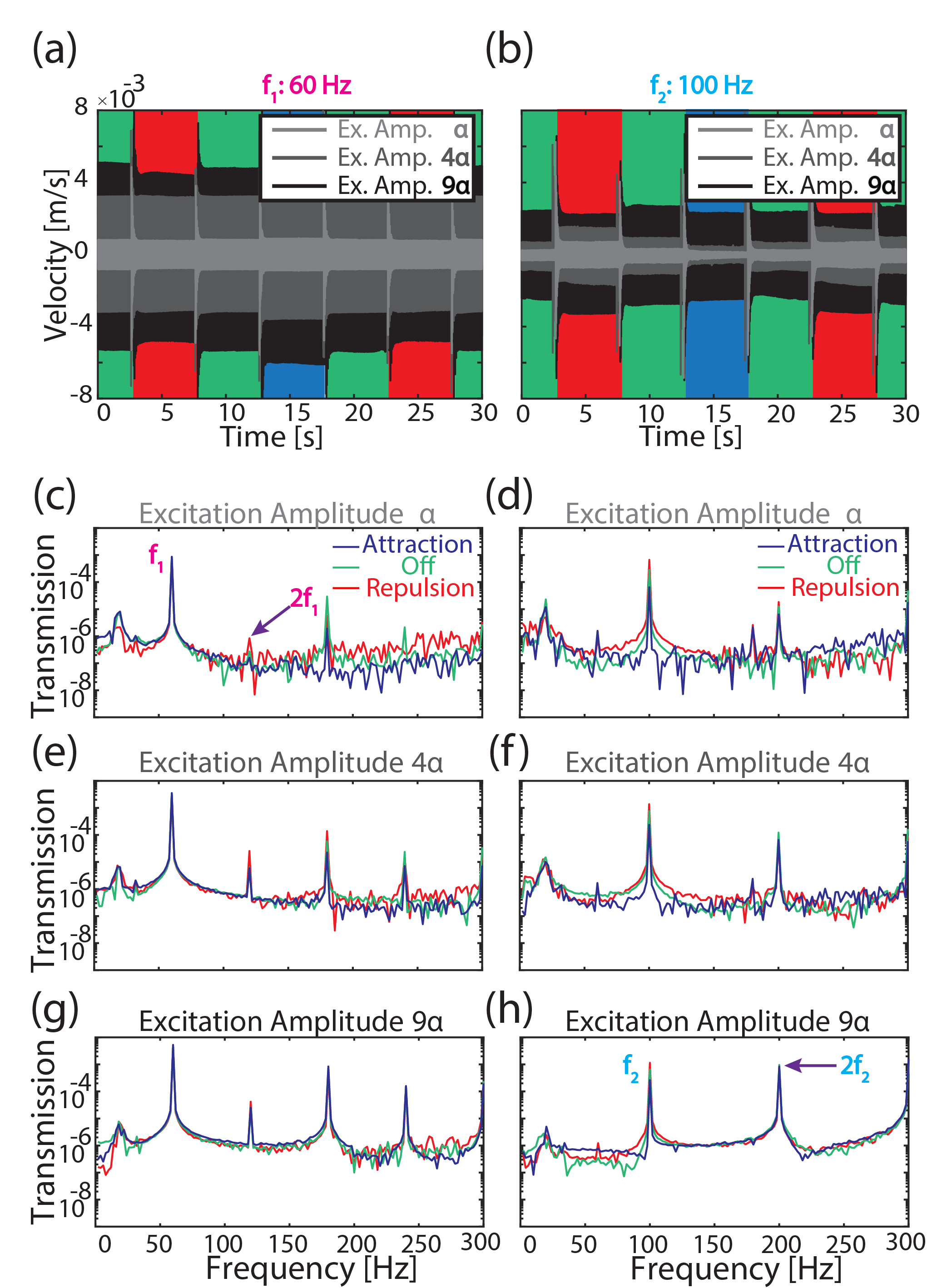}
\caption{\label{fig:Figure-4} \textbf{Programming metamaterial in time:} Time-signal of the selected frequencies from the  (a) pass band at $f_1=60 Hz$, (b) bandgap at $f_2=100 Hz$ (Under different excitation amplitudes: $\alpha$, $4\alpha$ and $9\alpha$). Transmitted signal’s fast Fourier transform (FFT) for (c,e,g) $f_1$ and (d,f,h) $f_2$ (Under different excitation amplitudes: $\alpha$, $4\alpha$ and $9\alpha$).}
\end{figure}

 In addition to the change in the amplitude of the transmitted signal at $f_2 = 100$ Hz, we observe a change in the transmitted waveform as a function of the magnetic field. To capture the nature of the change in the propagating wave in the time domain, we calculate  the transmitted signal's fast Fourier transform (FFT) at different magnetic-field states (Fig.\ref{fig:Figure-4}d). We note an order of magnitude reduction in the fundamental frequency amplitude between attractive and repulsive magnetic fields at $f_2 = 100$ Hz. The amplitude of the second harmonic, $2*f_2 = 200$ Hz, however, experiences less change because $2*f_2$ falls in a pass band frequency. The amplitude of the fundamental frequency at $f_1 = 60$ Hz is not affected by changes in the magnetic field because it is in a pass band; however, the second harmonic of $2*f 1 = 120$ Hz is a stop band frequency that is attenuated.  We increase the amplitude of the excitation from $\alpha$ to $9\alpha$, while shifting the magnetic field from attraction to Off to repulsion every 5 seconds. At both $f_1$ and $f_2$, we superimpose the transmitted time signal for the different excitation amplitudes (Fig.\ref{fig:Figure-4} (a-b)). At an excitation amplitude of $4\alpha$, the metamaterial can still effectively attenuate the excited wave with the magnetic field ON-attraction vs ON-repulsion. When the excitation amplitude approaches $9\alpha$, the transmitted signal's amplitude loses sensitivity to changes in the magnetic field, and the wave propagates at similar amplitudes at the band gap frequency $f_2 = 100$ Hz, whether the magnetic field is ON or OFF. The FFT of the transmitted wave at 100 Hz shows that the amplitudes of both the fundamental excitation frequency and its second harmonic are comparable. In other words, the metamaterial can block a second-harmonic-generated frequency, without affecting the fundamental frequency (Fig.\ref{fig:Figure-4}c,e, and g). Furthermore, with very large excitation amplitudes, the metamaterial can attenuate the fundamental frequency to a level below the generated second harmonic (Fig.\ref{fig:Figure-4}h).

In this study we propose an on-the-fly reprogrammable metamaterial that can control elastic waves without any moving parts. The metamaterial can be tuned in frequency and time, and the tuning is reversible. The study includes numerical and experimental analysis of the metamaterial, showing that the frequency bandgap can be shifted with a magnetic field and the transmission can be controlled by programming the metamaterial in time. Furthermore, the metamaterial can selectively attenuate certain spectral components beyond the excited fundamental frequency. Our metamaterial could be utilized in many potential applications in advanced acoustic devices.

\paragraph*{Methods} We fabricate the metamaterial using a laser cuter (Full spectrum, pro-series-PS 24). The lattice constant of the fabricated metamaterial is $a=22 mm$. The metamaterials are fabricated out of acrylic sheets with Young’s modulus $E=3.2026E+9 Pa$, density $\rho=1180 kg/m^3$,  and Poisson’s ratio  $\nu=0.35$. The fabricated unit cells have asymmetric profile along their thickness. The spiral cutting width, $w$, varies from top ($w=0.6 mm$) to bottom ($w=0.3 mm$) with average cutting width of $\approx 0.45 mm$. The embedded magnets at the center of each unit cell are cylindrical neodymium magnets with a $3 mm$ diameter.  An array of electromagnets (type: kk-P20/25, DC12V Kaka Electric) was placed underneath the metamaterial sample. All the electromagnets were connected in parallel to an Arduino circuit powered by a DC supply (type: LW PS-3010DF). The distance between the metamaterial sample and the electromagnets in the OFF state is $5 mm$. We excite the metamaterial with a mechanical shaker (type: Brüel \& Kjær 4180) at one end, while the other end is fixed. We measure the response of the metamaterial by a scanning laser Doppler vibrometer (Polytech-PSV-500).


%

\newpage
\beginsupplement

\begin{widetext}
\newpage\hspace{-3mm}\Large{\textbf{Supporting Information: \\}}
\Large{\textbf{Harnessing asymmetry to  reprogram nonlinear metamaterials on-the-fly with no moving parts}\\}

\section{Unit cell analysis}

The unit cell is composed of  two symmetric cuts in the shape of concentric Archimedean spirals. The equation of the   spiral can be written   in polar coordinates \cite{foehr2018spiral} as $r(s) = R - (R-r)s, ~\phi(s) = 2\pi ns$, where $r=0.25a$ is the inner radius, $R=0.475a$ is the outer radius,  $n=2$ is the number of turns, $a=22 mm$ is the lattice constant, $w=0.45 mm$ is the cutting width of spirals (average of both bottom and top cutting width), and $p=3 mm$ is a diameter of the permanent magnet (Fig.\ref{fig:SFigure-1} a-b). We fabricate the metamaterial using laser cutting. Despite the symmetric design of the unit cell, the spiraling cuts have asymmetric profiles in the thickness direction, meaning the cutting width of the spirals varies from the top to the bottom surface of the metamaterial(Fig.\ref{fig:SFigure-1} c)

\begin{figure}[!h]
\centering
\includegraphics{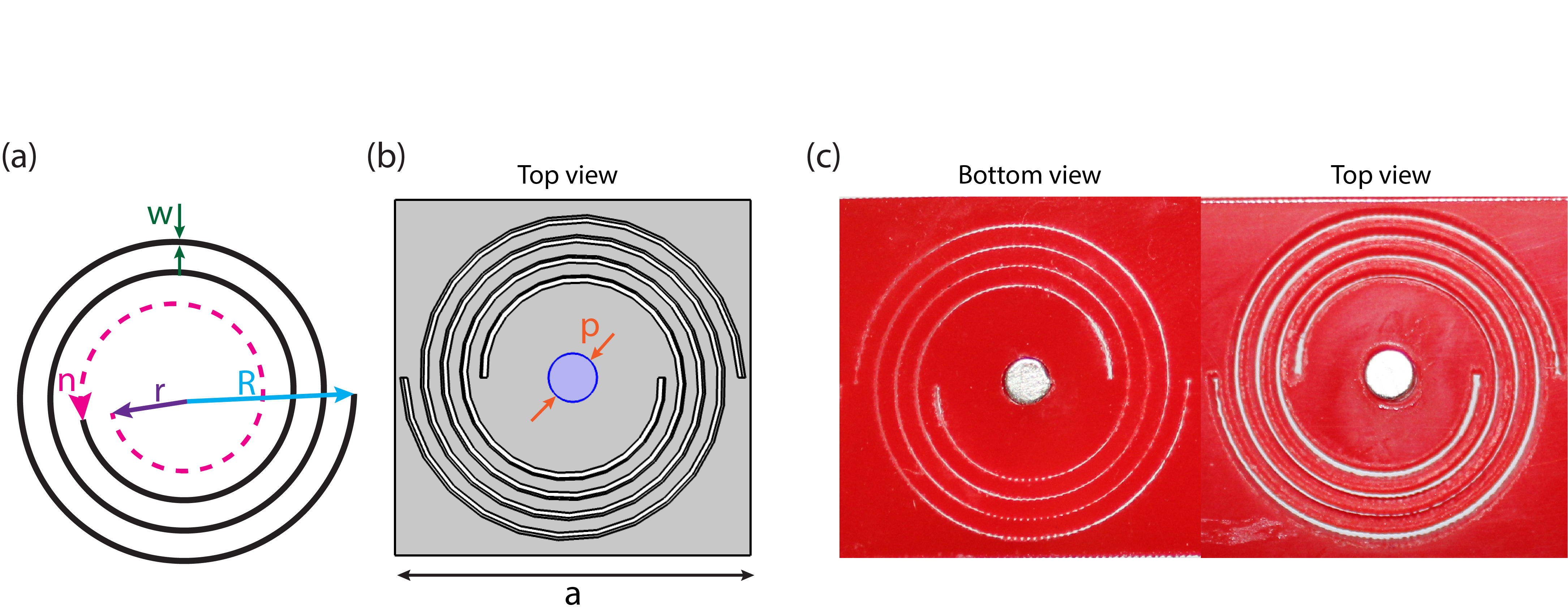}
\caption{\label{fig:SFigure-1}\textbf{Unit cell design:} (a) Geometric parameters of an Archimedean spiral. (b) Designed unit cell. (c) Bottom and top view of the fabricated unit cell.}
\end{figure}
The position of the band gap (BG) within the frequency spectrum is a function of the unit cell design parameters, including the cutting width of the spiral. To study the influence of the unit cell parameters on the BG, we consider a symmetric unit cell with different cutting width, while keeping all other parameters fixed. We compute the dispersion curves for three unit cell with cutting width $w = 0.3, 0.45$, and $0.6~mm$. Indeed, as the cutting width increases the BG frequency decreases. Such decrease in frequency is a result of the reduction of the effective stiffness of  the unit cell core. To confirm our infinite model calculations, we numerically simulate a finite structure made out of 11 unit cells. We excite the finite structure at one end and record its transmission at the other end. The resulting frequency response functions show a clear shift in the attenuated frequency ranges as predicted by the infinite unit cell model (Fig.S2 a-c). 
\begin{figure}
\centering
\includegraphics{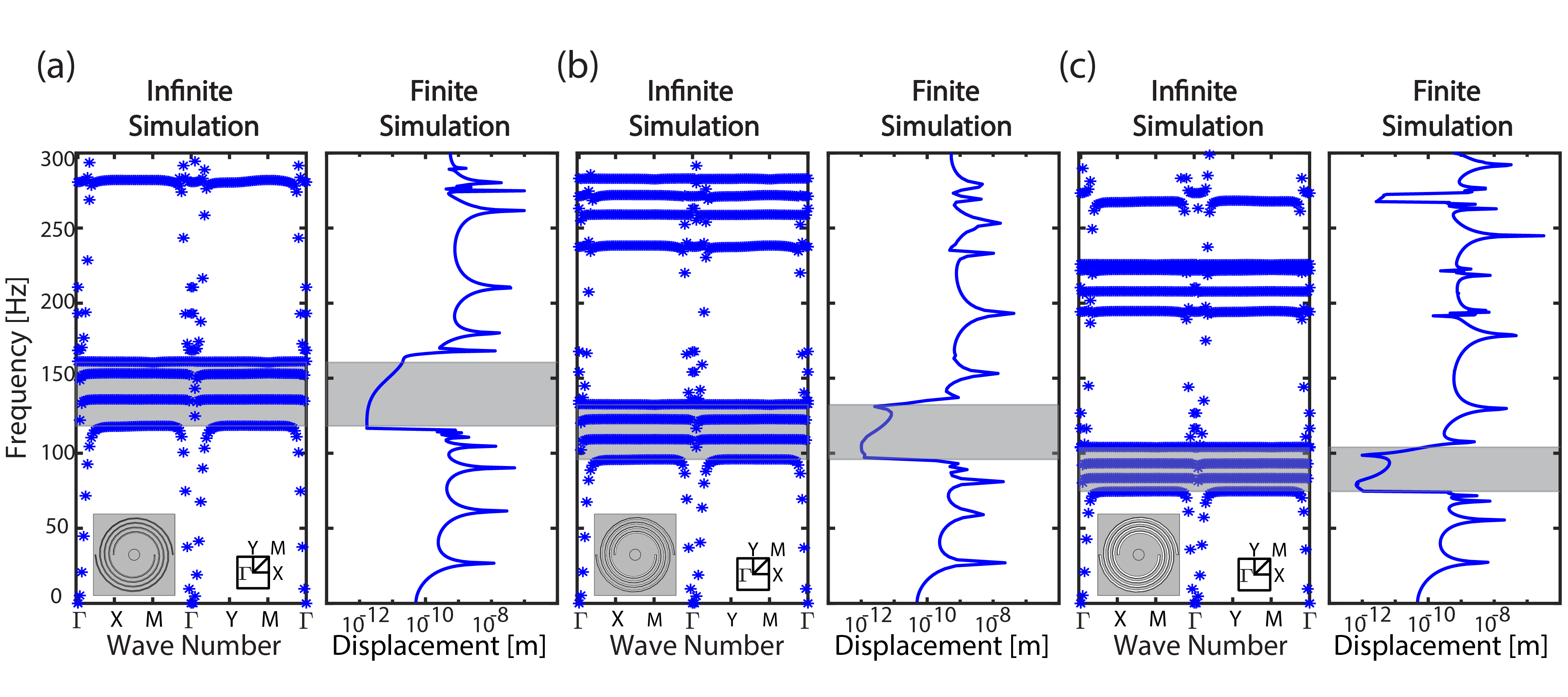}
\caption{\label{fig:SFigure-2}\textbf{Unit cell design parameters:} Numerically computed dispersion curves for infinitely repeated symmetric unit cells and numerically computed frequency response function of the transmission in metamaterial having an array of unit cells of size 11x1. Each unit cell composed of two concentric Archimedean spirals having the cutting width of (a) $w = 0.3 mm$, (b) $w = 0.45 mm$, and (c) $w = 0.6 mm$. (band gap regions are highlighted in gray)}
\end{figure}

\section{symmetric vs asymmetric unit cell analysis}

Next, we consider the influence of unit cell symmetry on the properties of our metamaterials. We start by comparing the dispersion curves of a symmetric and an asymmetric unit cells. For the symmetric unit cell, the spiral cut has a width of $w = 0.45 mm$, while the asymmetric unit cell has top surface cut of ($w_t = 0.6 mm$) and bottom surface cut ($w_b = 0.3 mm$). We choose $w_{t}$ and $w_b$ such that the average is equivalent to the symmetric unit cell cutting width (Fig.S3 c). The resulting band structures show very similar behavior at low frequency, however as the frequency increases, the two band diagrams start to deviate from each other as more complex modes start to appear   (Fig.S3 a). To confirm the band structure calculations, we consider a finite structure analysis for both unit cells and compute the frequency response of the metamaterials. At low frequencies, the difference in the frequency response of the two different unit cells is indeed negligible, but as the frequency increases, the difference between the two becomes more clear, in agreement with our band structure analysis (Fig.S3 b).

\begin{figure}[!h]
\centering
\includegraphics{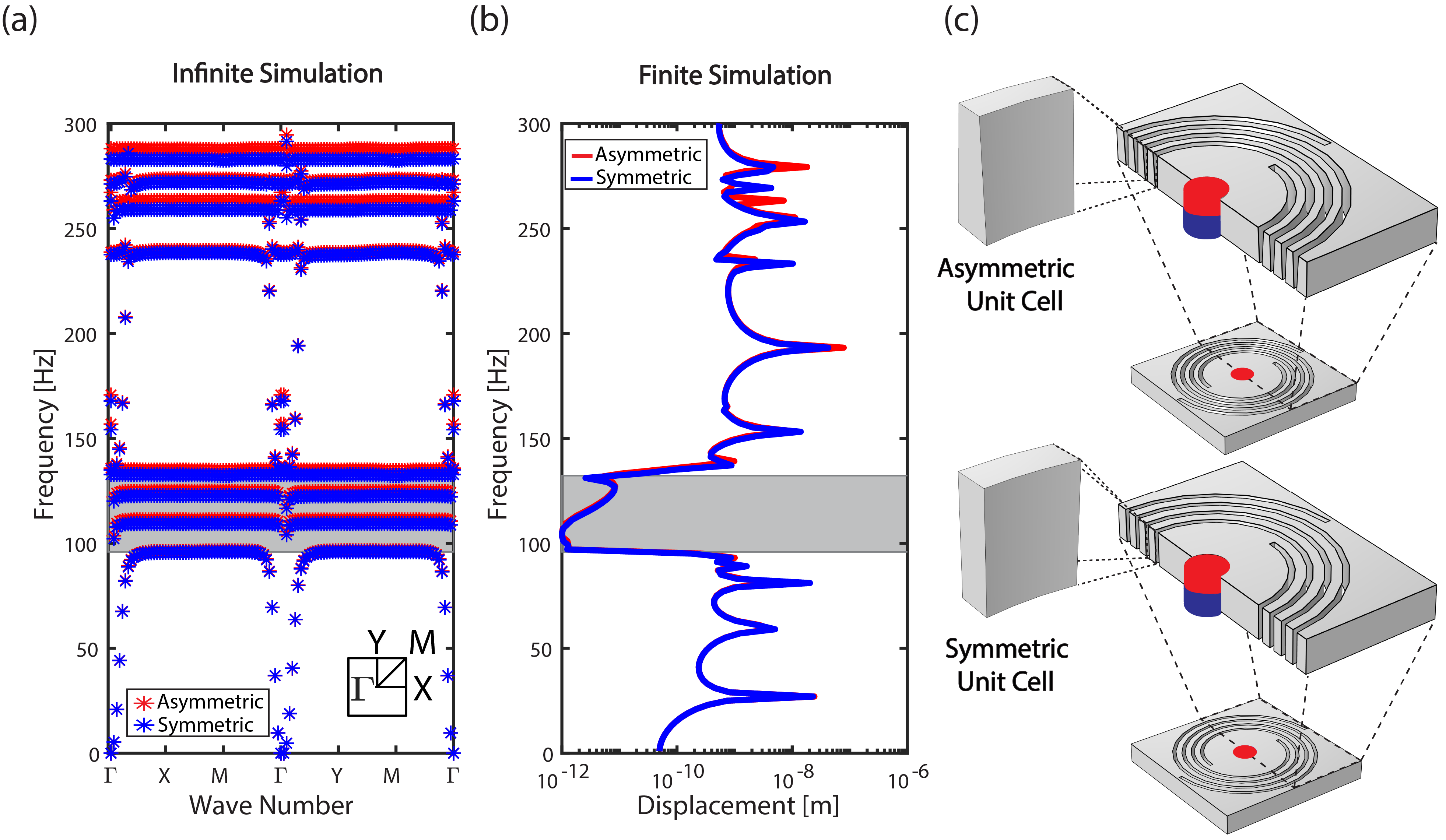}
\caption{\label{fig:SFigure-3}\textbf{Asymmetric and symmetric unit cell analysis:} (a) Dispersion curves of both symmetric and asymmetric unit cells. (b) Frequency response of the metamaterial composed of the two different unit cells. (c) Asymmetric and symmetric unit cells.}
\end{figure}

\section{One-dimensional vs two-dimensional periodicity}

For completeness, we also examine the effect of different types of periodicity on the  unit cell band structure. We consider an infinite model with periodicity along only the x-direction (rather than two-direction periodicity along both x and y-directions). In such scenario, we can only calculate the dispersion curve along the $\Gamma-X$ path of the irreducible Brillouin zone. The dispersion curves show a band gap (a gray highlighted region in (Fig.S4 a)) that matches the band gap of the two-directions periodic unit cell (Fig.2 (a)). Both mode shapes at the edge of the band gap at 96 Hz and 135 Hz correspond to out-of-plane displacement profiles of the center of the unit cell. Within the  band gap frequency range, we also have there transmission branches at 109, 123, and 132 Hz which correspond to in-plane modes (Fig.S4 b).

\begin{figure}
\centering
\includegraphics{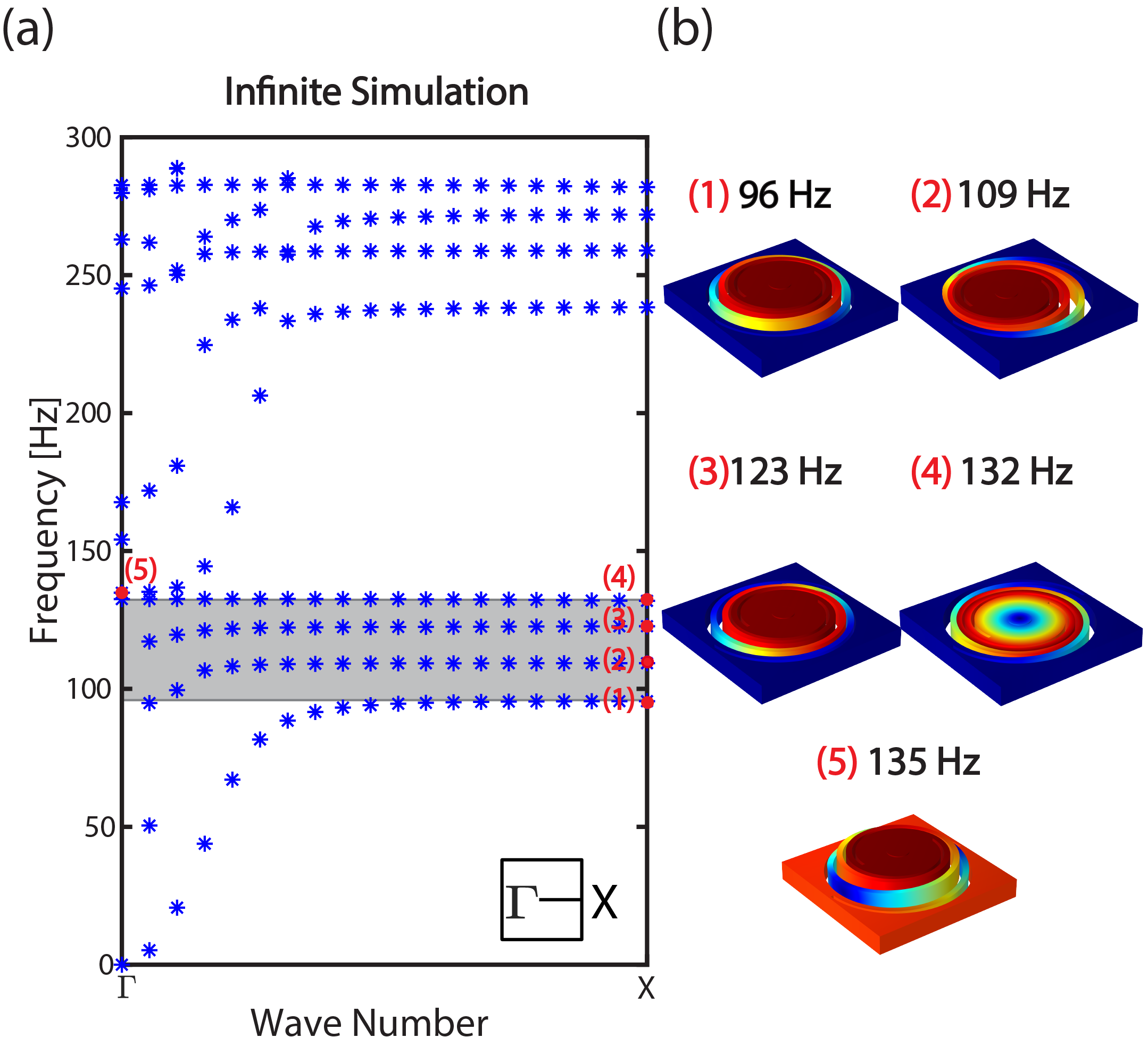}
\caption{\label{fig:SFigure-4}\textbf{1D unit cell analysis :} (a) Dispersion curves of the 1D symmetric unit cell. (b) Mode shapes (the mode shapes at the
edges of the band gap correspond to out-of-plane displacement profiles for the center of the unit cell at 96 Hz and 135 and another three transmission branches at 109, 123, and 132 Hz correspond to in-plane displacement) for the spirals. (band gap regions are highlighted in gray)}
\end{figure}

\begin{figure}
\centering
\includegraphics{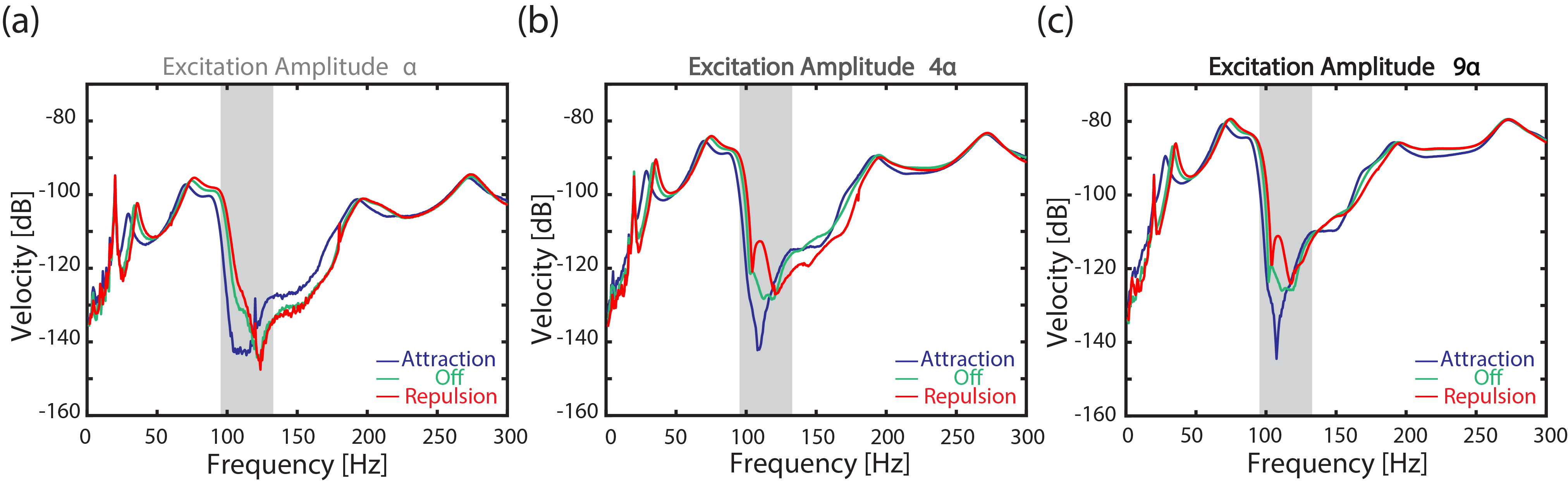}
\caption{\label{fig:SFigure-5}\textbf{Programming metamaterial in frequency:} Experimentally measured frequency response of the proposed metamaterial under excitation amplitude of (a) $\alpha$, (b) $4\alpha$ and (c) $9\alpha$. (band gap regions are highlighted in gray)}
\end{figure}

\section{Programming metamaterial in frequency}

We examine the response of our metamaterial in the frequency domain under different excitation amplitudes $\alpha$, $4\alpha$, and $9\alpha$. We start by harmonically exciting one end of the metamaterial sample in the presence of two different magnetic-field polarizations and record the amplitude of the transmitted elastic waves at the other end. At excitation amplitude $\alpha$, we observe a clear shift of the band gap frequency from attraction to repulsion (Fig.S5 a). By increasing the amplitude of the harmonic excitation from $\alpha$ to $9\alpha$, we still see a shift of the band gap frequency due to the presence of the different magnetic fields, however, the amplitude of the attenuation within the band gap decreases for OFF and ON-repulsion cases (Fig.S5 b-c). 

\section{Programming metamaterial in time}

\begin{figure}[!h]
\centering
\includegraphics{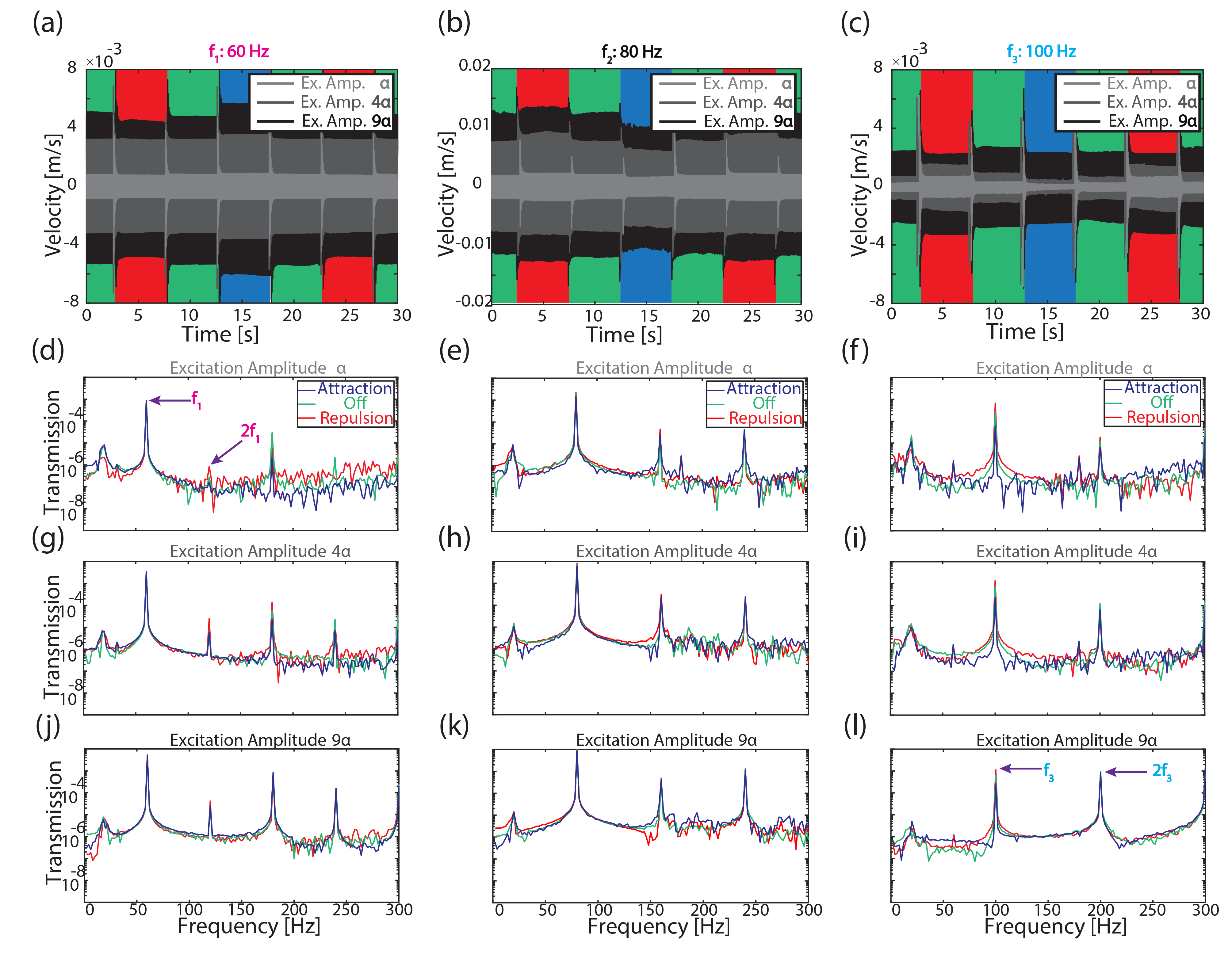}
\caption{\label{fig:SFigure-6}\textbf{Programming metamaterial in time:} Time-signal of the selected frequencies from the pass band at (a) $f_1=60$ Hz, (b) $f_2=80$ Hz, and band gap at (c) $f_3=100$ Hz (under different excitation amplitude of $\alpha$, $4\alpha$ and $9\alpha$). Transmitted signal’s fast Fourier transform (FFT) for (d-g-j) $f_1$, (e-h-k) $f_2$, and (f-i-l) $f_3$ (under different excitation amplitude of $\alpha$, $4\alpha$ and $9\alpha$).}
\end{figure}

\begin{figure}[b]
\centering
\includegraphics{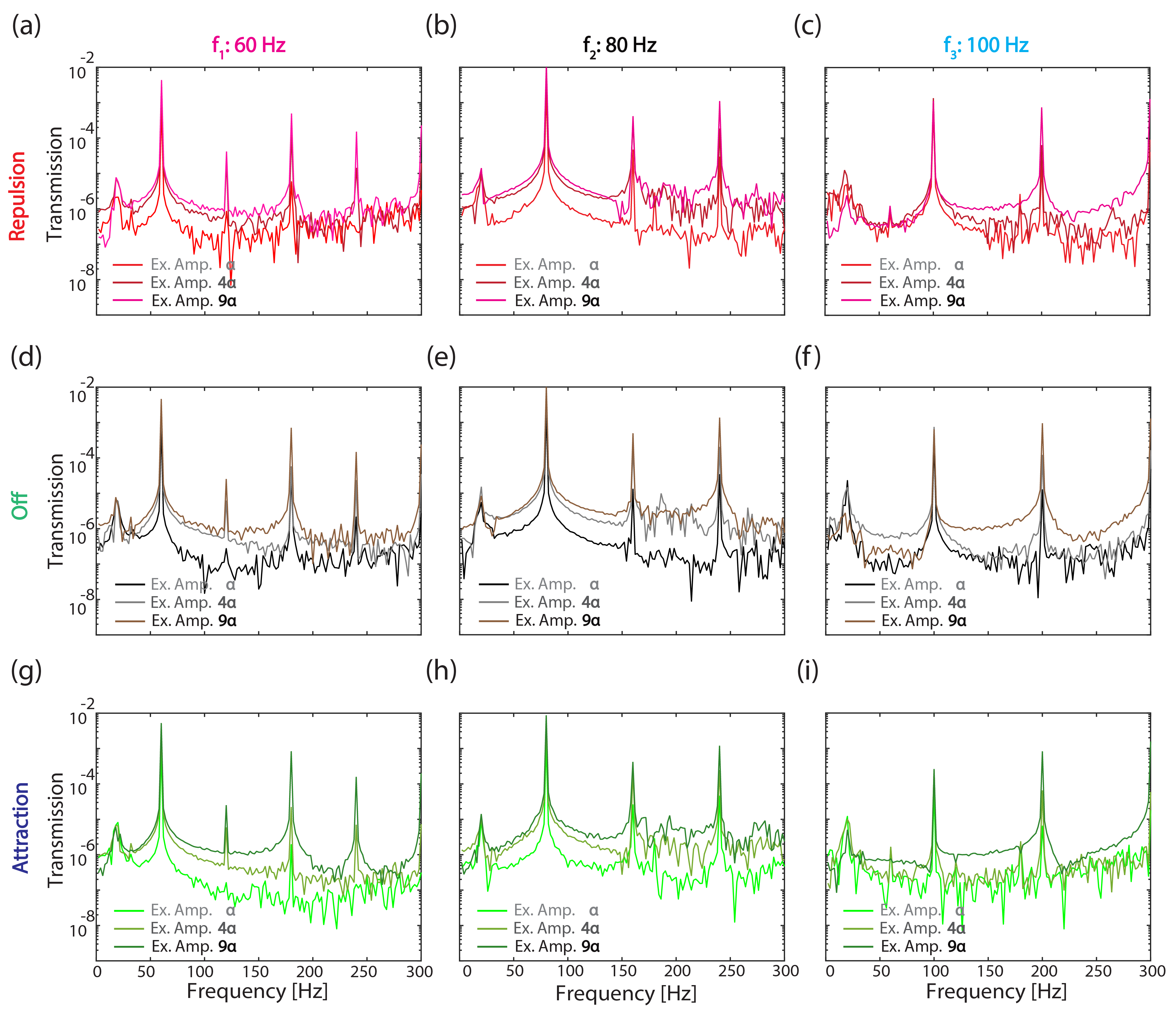}
\caption{\label{fig:SFigure-7}\textbf{FFT comparison:} Comparing the
amplitude of the transmitted wave at 60, 100, and 80 Hz for (a-c) ON-repulsion, (d-f) OFF and (g-i) ON-attraction under different excitation amplitude of $\alpha$, $4\alpha$ and $9\alpha$}
\end{figure}

After characterizing the behavior of the metamaterial in the frequency domain we examine the proposed metamaterial in the time domain. First, we program the magnetic field to change between OFF, ON-repulsion, and ON-attraction. Next, we excite one end of the sample with a harmonic excitation and record the transmitted elastic wave through the metamaterial at the other end for two  pass band frequencies ($f_1 = 60$ Hz and $f_2 = 80$ Hz) and one band gap frequency ($f_3 = 100$ Hz) (Fig.S6 a-c). At frequencies $f_1 = 60$ Hz and $f_2 = 80$ Hz, the transmission amplitude is almost the same between OFF, ON-repulsion, and ON-attraction because $f_1$ and $f_2$ are pass band frequencies. At the frequency $f3 = 100$, the transmission amplitude changes by changing the magnetic field between OFF, ON-repulsion, and ON-attraction. For example, the transmission amplitude is 6 times higher in the case of repulsion than attraction. We calculate the transmitted signal’s fast Fourier transform (FFT) for all frequencies.  Results show that the amplitude of the fundamental frequencies at $f_1 = 60$ Hz and $f_2 = 80$ Hz are not affected by changes in the magnetic field from OFF, ON-repulsion to ON-attraction (Fig.S6 d,e). The reason is that both frequencies are in the pass band; however, there is a change at the second harmonic of $2*f_1 = 120$ Hz because it falls in the band gap frequency range (Fig.S6 d). But at the second harmonic of $2*f_2 = 160$ Hz the wave is transmitted because it falls in the pass band frequency (Fig.S6 e). Also, we observe that the amplitude of the second harmonic, $2*f_3 = 200$ Hz shows less change and the reason is that $2*f_3$ falls in a pass band frequency range (Fig.S6 f). Afterwards, we increase the amplitude of the excitation from $\alpha$ to $4\alpha$, while still changing the magnetic field polarization every 5 seconds. At $2*f_3$ the  metamaterial can still attenuate the transmitted wave (Fig.S6 i). However,  when we increase the  amplitude of the excitation from $4\alpha$ to  $9\alpha$, the transmitted amplitude is not affected by the change in the magnetic field and the wave propagates at similar amplitudes at the band gap frequency $f_3$ (Fig.S6 l). It should be noted that when the  excitation amplitude is $9\alpha$, the transmitted amplitude at $3*f_2 = 240$ Hz is higher than that of $2*f_2 = 160$ Hz, which means that the wave is transmitted at higher amplitude at $3*f_2$ than the second harmonic peak (Fig.S6 k). For better comparing the amplitude of the transmitted wave in the frequency domain, we separate the FFTs based on the magnetic polarizations (Fig.S7).

\section{Numerical simulation of the magnetic field}
\begin{figure}[b]
\centering
\includegraphics{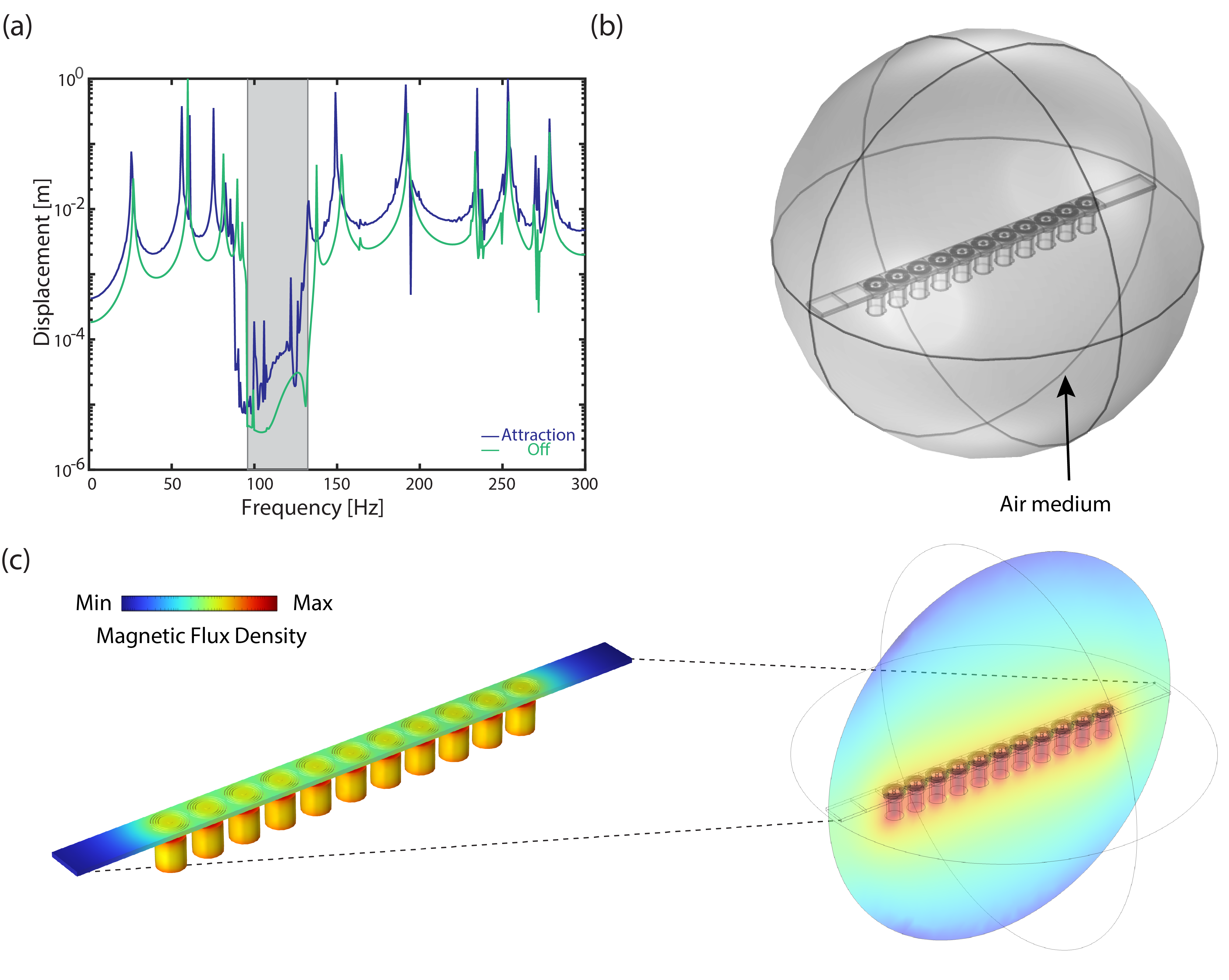}
\caption{\label{fig:SFigure-8}\textbf{Numerical simulation of the magnetic field:} (a) Frequency response of the metamaterials under magnetic field (band gap region is highlighted in gray). (b) Simulated model of the proposed metamaterial having electromagnets to create a magnetic field over a sample which covered by air medium. (c) Magnetic flux density.}
\end{figure}

To examine the metamaterial in the frequency domain under the influence of the magnetic field numerically, we utilize COMSOL multi-physics version 6.0. To compute the frequency response function of the transmission in the metamaterial considering the magnetic field effect we use the AC/DC module (magnetic field physics). We consider an array of 11 symmetric unit cells with fixed-fixed boundary conditions. In addition, we model 11 electromagnets and place them underneath the metamaterial sample and surround the sample and electromagnets with an air medium (Fig.S8 b). We apply an out-of-plane harmonic excitation at one end of the metamaterial and record the transmission of the wave at the other end of the metamaterial. We compute the frequency response function of the transmission in the metamaterial for OFF and ON-attraction cases (Fig.S8 a), and the computed magnetic flux density within the medium (Fig.S8 c). 

\section{ Programming time window}
\begin{figure}[b]
\centering
\includegraphics{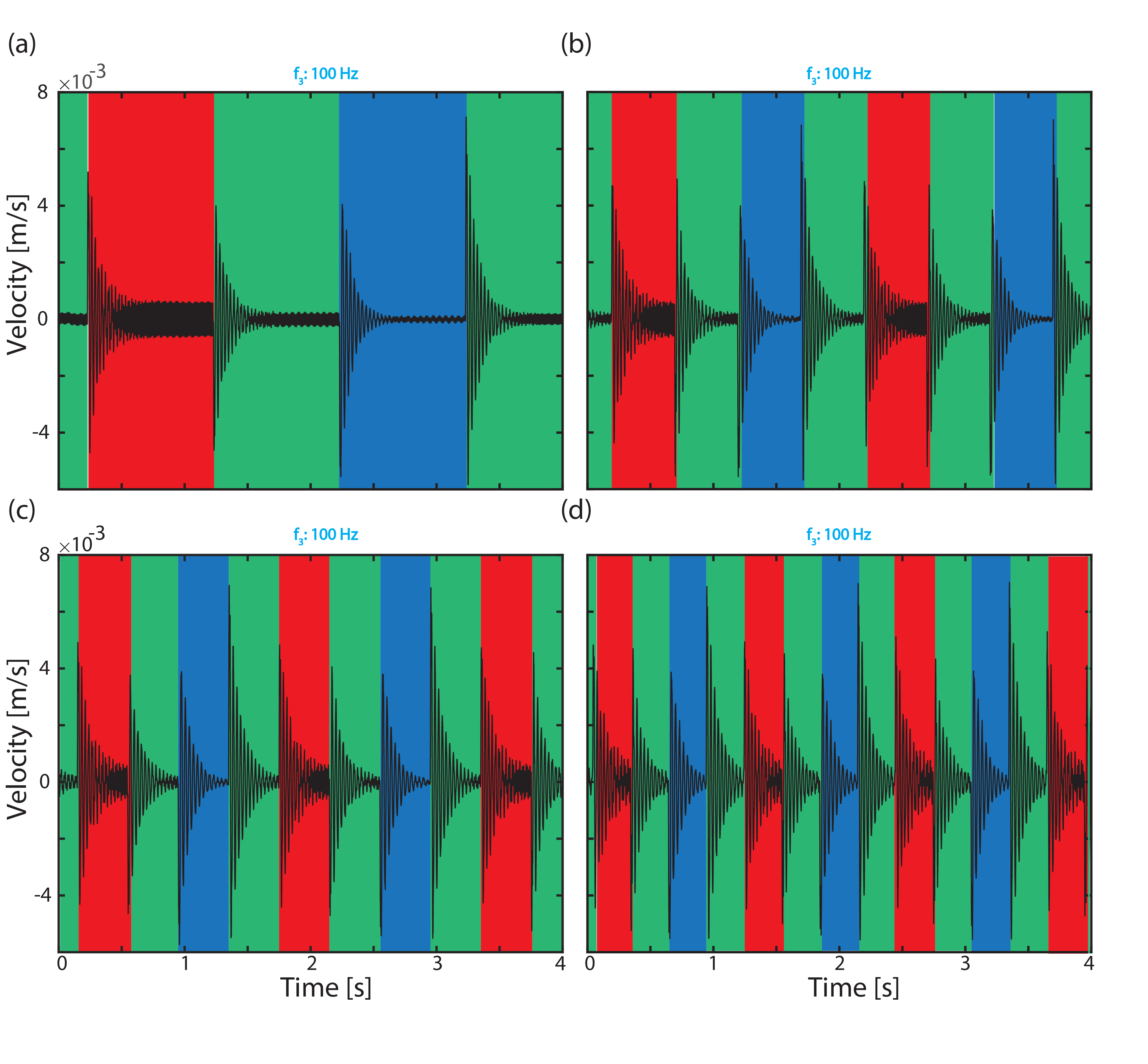}
\caption{\label{fig:SFigure-9}\textbf{Programming time window:} Program the magnetic field to change between OFF, ON-repulsion, and ON-attraction at (a) 1000 ms, (b) 500 ms, (c) 400 ms and (d) 300 ms, for the selected frequency from the band gap at $f_3=100 Hz$ considering excitation amplitude of $\alpha$.}
\end{figure}
To test the limits of our programming in time, we program the magnetic field to change between OFF, ON-repulsion, and ON-attraction at different time windows (Fig.  \ref{fig:SFigure-9}). The excitation frequency is fixed at $f_3 = 100$ with an excitation amplitude of $\alpha$. The change in magnetic field happens at windows of 1, 0.5, 0.4 and 0.3 seconds.

\section{ Measured static and magnetic forces}

\begin{figure}[!h]
\centering
\includegraphics{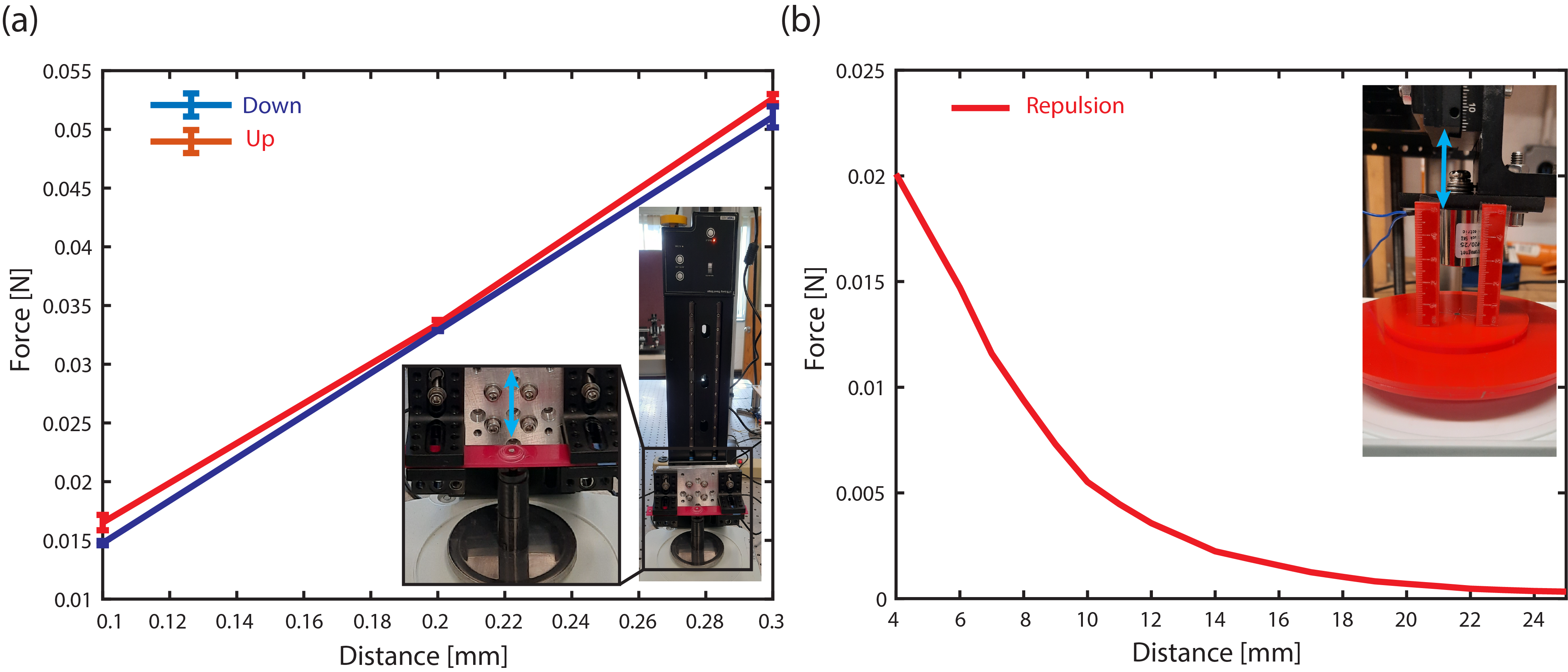}
\caption{\label{fig:SFigure-10}\textbf{Measured static and magnetic force:} (a) The measured static force. The inset shows an experimental setup where a linear translation stage (Thorlabs, LTS-300) is used to move the unit cell up and down to measure the static force. (b) The measured repulsion force between the magnet and the electromagnet. The inset shows the experimental setup where the electromagnet is mounted on a stage to move toward a magnet direction to measure the force-distance curve. (Light blue double-headed arrow shows the direction of the displacement)}
\end{figure}

\begin{figure}[!h]
	\centering
	\includegraphics{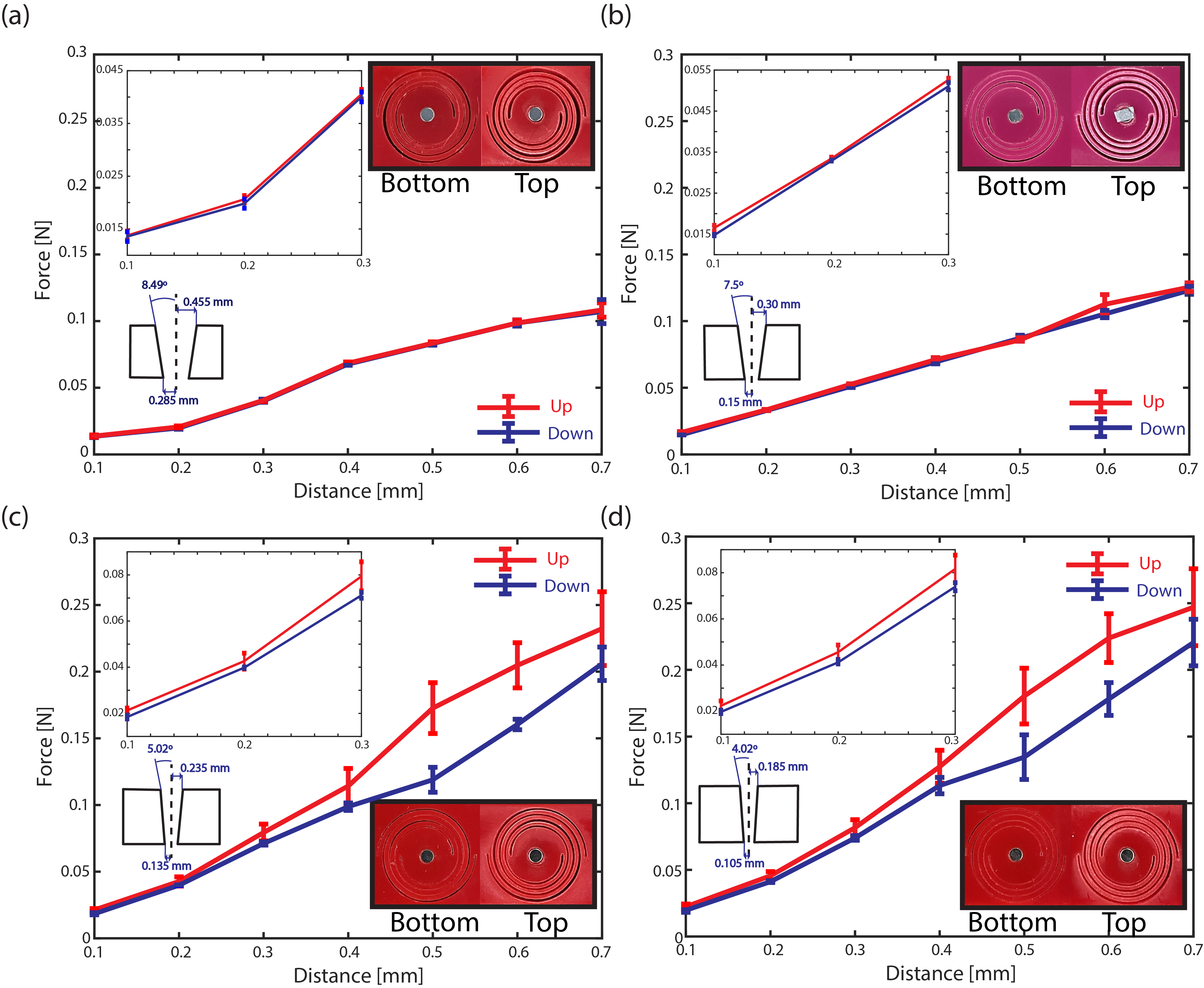}
	\caption{\label{fig:SFigure-11}\textbf{Measured static force:}  Measured static forces for different unit cells with asymmetric profiles. The plane angles of the cuts are: (a) 8.49$^\circ$, (b) 7.5$^\circ$, (c) 5.02$^\circ$, (d) 4.02$^\circ$. The insets on the right side of the plots show the top and bottom surface of each fabricated unit cell, and the insets on the left side show a schematic  side-view of each asymmetric profile with its plane angle, and the varying cutting width $w$ from the top to the bottom of each unit cell.}
\end{figure}

To measure the force, we fabricate a single-unit cell and mount it on a linear translation stage (Thorlabs, LTS-300 Fig.S10 a-inset). We move the sample by an increment of 0.1 mm to measure the force at each step through a load cell (Fig.S10 a). The measured force in both cases shows that the spiral-spring is stiffer when its center goes up and than when its center goes down. This stiffness difference is due to the presence of asymmetrical profiles in the unit cell. Meaning, the cutting width of the spirals varies from the top to the bottom surface of the metamaterial sample as proved by static force measurement. 

Moreover, we also investigate the asymmetric unit cell profiles and its correlation to plane angles.  To do this, we fabricate three more samples with different width of spiral cuts, which varies from top to bottom surface. To measure the angles, we consider the top edge of the surface to the plane (Fig.S11 insets). The results shown in (Fig.S11) are the measured force for the first unit cell with the spiral cutting width, where $w$ varies from  top ($w=0.91 mm$) to bottom ($w=0.57 mm$), having the plane angle 8.49$^\circ$ (Fig.S11 a). The second unit cell with the spiral cutting width, $w$, varies from top ($w=0.47 mm$) to bottom ($w=0.27 mm$) with the plane angle 5.02$^\circ$ (Fig.S11 c). The third unit cell with the spiral cutting width, $w$, varies from top ($w=0.37 mm$) to bottom ($w=0.21 mm$) having the plane angle 4.02$^\circ$ (Fig.S11 d). Also, the plane angle for the primary unit cell (Fig.S10 a) and (Fig.S11 b) is 7.5$^\circ$. The results show the stiffness differences due to the presence of asymmetrical profiles in the unit cell. 

In addition, we characterize the magnetic force between an electromagnetic and a permanent magnet. We fabricate a single unit cell with an embedded magnet at its center. We start incrementally  move the electromagnet closer to the magnet with a linear stage from  $25mm $ to $4 mm$ (Fig.S10 b).

\section{Metamaterial quality factor}
We characterize the quality factor of the material that we use to fabricate the metamaterials, by measuring the resonance amplitude of the sample using a mechanical shaker as a harmonic excitation source. We select the first resonance peak within the frequency response plot (Fig. S12). We calculate $Q=f/\Delta_f $ which is $\approx$50, where $f$ is the resonance frequency and $\Delta_f $ is the width of the frequency range.

\begin{figure}[!h]
\centering
\includegraphics[scale = 0.75]{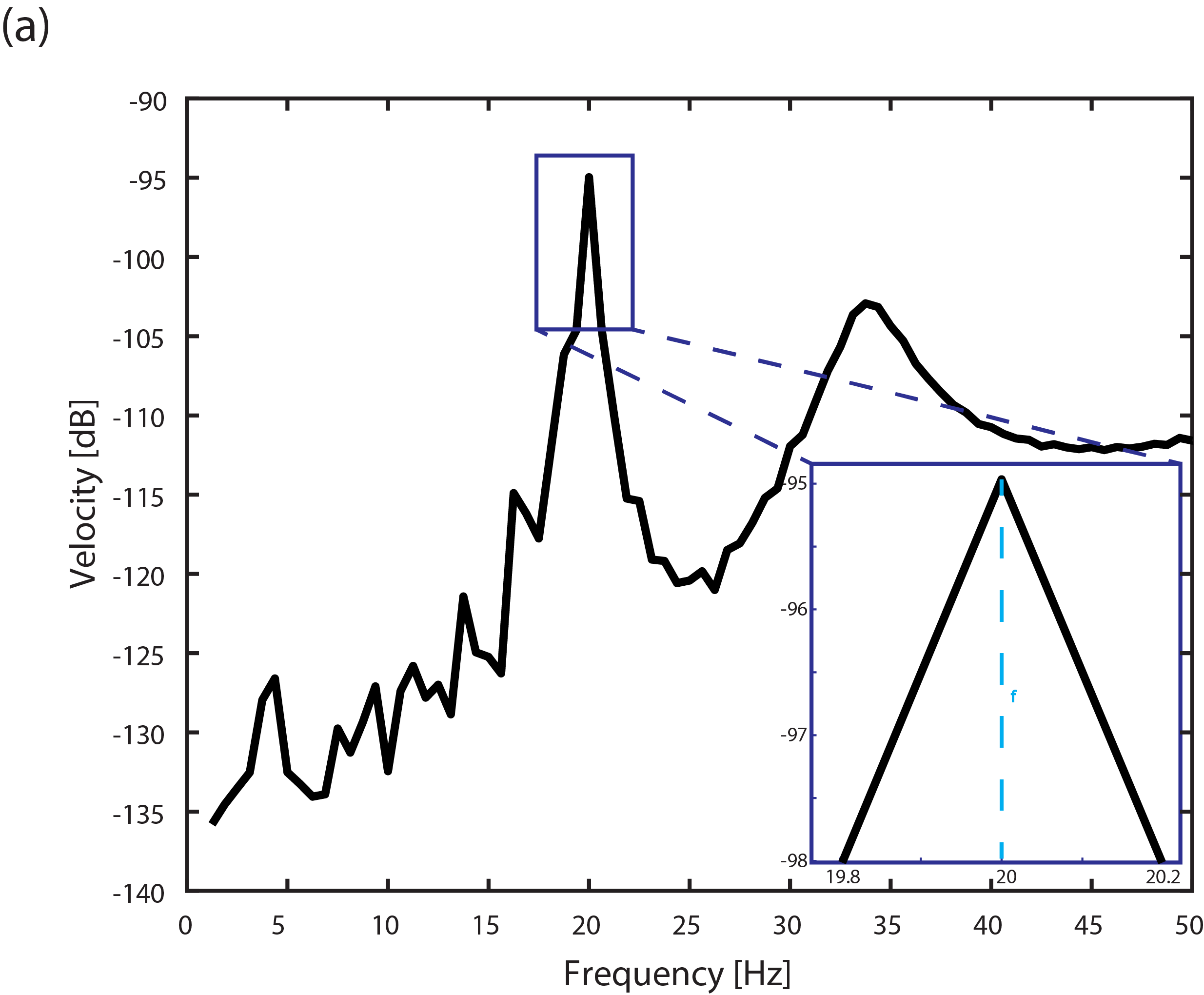}
\caption{\label{fig:SFigure-11}\textbf{Quality factor of the metamaterial:}  (a)  First selected resonance peak within the experimentally measured frequency response of the proposed metamaterial under excitation amplitude of $\alpha$}
\end{figure}

\section{Force-displacement simulation}

\begin{figure}
\centering
\includegraphics{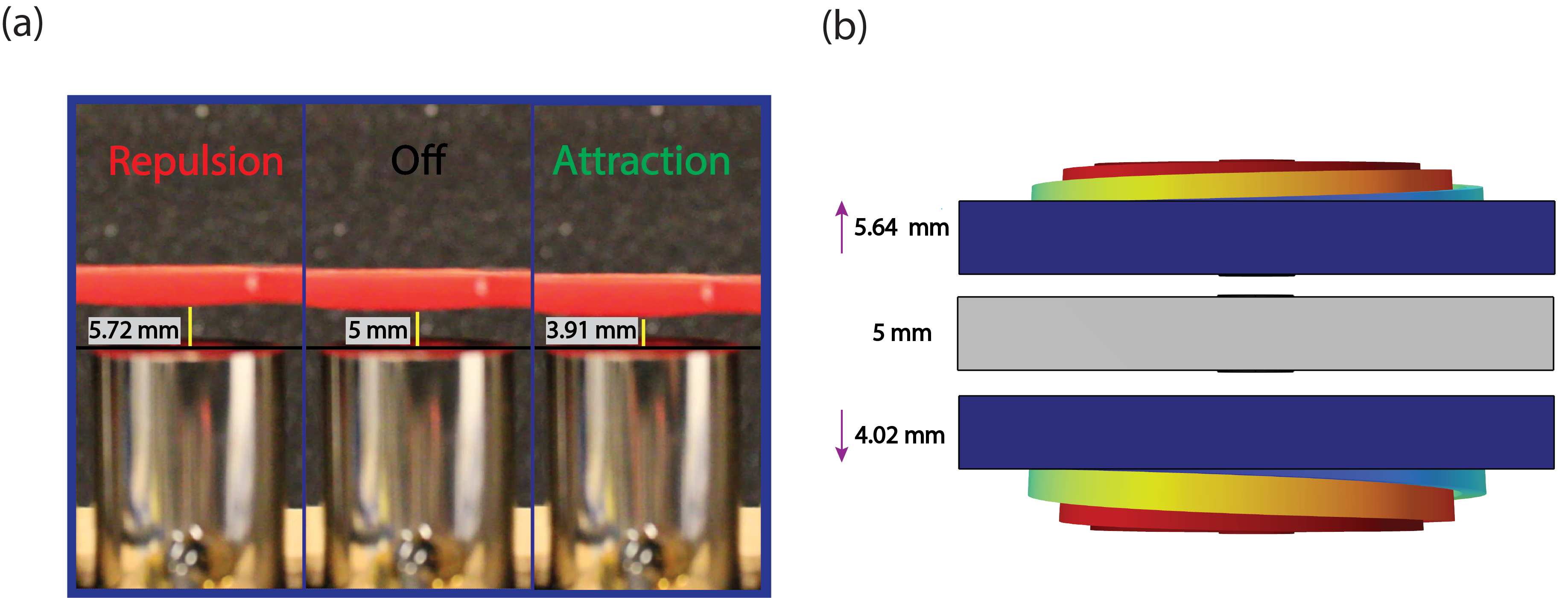}
\caption{\label{fig:SFigure-12}\textbf{Force-displacement evaluation:}  (a) Change in the displacement between OFF, ON-repulsion, and ON-attraction, b) numerical force-displacement mode shapes.}
\end{figure}

We experimentally measure the distance between the electromagnet and the unit cell for each of the cases (1) magnetic field OFF, (2) ON-repulsion, and (3) ON-attraction. The distance when the magnetic field is OFF = $5~mm$. The displacement for the repulsion ($+0.72~mm$) and for attraction is ($-1.09~mm$) (Fig. S12 a). We utilize the experimentally obtained values for the reaction forces at the center of the spiral to validate our numerical model. We calculate the displacement of $0.64~mm$ for the case when we apply the force in the up direction (center of spirals goes up) and the displacement of $0.98~mm$ for the case when we apply the force in the down direction (center of spirals goes down). There is a very good agreement between the experimentally obtained static force-displacement values and the numerical ones (Fig. S13 b). 


\textbf{Supplementary movie 1} The movie shows the transmission of the wave through the  re-programmable phononic metamaterial at pass band $f_1 = 60$ Hz.

\textbf{Supplementary movie 2} The movie shows the transmission of the wave through the  re-programmable phononic metamaterial at pass band $f_2 = 80$ Hz.

\textbf{Supplementary movie 3} {The movie shows the attenuation of the wave through the  re-programmable phononic metamaterial at band gap $f_3 = 100$ Hz.}

%
%
%
%

\newpage
\end{widetext}

\end{document}